\documentclass[12pt,twoside,fleqn,espcrc1]{article}
\usepackage{espcrc1}
\usepackage{epsfig}
\usepackage{graphics}
\psfigdriver{dvips}

\begin{document}

\vspace*{1.5cm} \noindent {\large  Hadron yields and spectra in 
Au+Au collisions at the AGS} \\ \ \\

\noindent {Roger Lacasse} \\ 

\noindent Foster Radiation Laboratory,
 McGill University, Montr\'eal, Canada H3A 2B1 \\

\noindent For the E877 Collaboration:

\noindent
J.~Barrette$^4$, R.~Bellwied$^{8}$, S.~Bennett$^{8}$, R.~Bersch$^5$,
P.~Braun-Munzinger$^5$, W.~C.~Chang$^5$, W.~E.~Cleland$^6$,
M.~Clemen$^6$, J. D.~Cole$^3$, T.~M.~Cormier$^{8}$, Y.~Dai$^4$, G.~David$^1$,
J.~Dee$^5$, O.~Dietzsch$^7$, M.W.~Drigert$^3$, K.~Filimonov$^4$,
J.~R.~Hall$^8$, T.~K.~Hemmick$^5$, N.~Herrmann$^2$, B.~Hong$^5$,
C.~L.~Jiang$^5$, S.~C.~Johnson$^5$, Y.~Kwon$^5$, R.~Lacasse$^4$,
A.~Lukaszew$^8$, Q.~Li$^{8}$, T.~W.~Ludlam$^1$, S.~McCorkle$^1$,
S.~K.~Mark$^4$, R.~Matheus$^{8}$, D.~Mi\'skowiec$^5$, 
J.~Murgatroyd$^8$, E.~O'Brien$^1$, 
S.~Panitkin$^5$, P.~Paul$^5$,
T.~Piazza$^5$, M.~Pollack$^5$, C.~Pruneau$^8$, Y.~Qi$^4$, M.~N.~Rao$^5$,
M.~Rosati$^1$, N.~C.~daSilva$^7$, S.~Sedykh$^5$, U.~Sonnadara$^6$,
J.~Stachel$^5$, N.~Starinski$^4$, E.~M.~Takagui$^7$, M.~Trzaska$^5$, S.~Voloshin$^6$,
T.~Vongpaseuth$^5$, G.~Wang$^4$, J.~P.~Wessels$^5$, C.~L.~Woody$^1$,
N.~Xu$^5$, Y.~Zhang$^5$, Z.~Zhang$^6$, C.~Zou$^5$
\\ \ \\
\noindent $^1$BNL -
$^2$GSI -
$^3$Idaho Nat. Eng. Lab. -
$^4$McGill Univ. -
$^5$SUNY, Stony Brook -
$^6$Univ. of Pittsburgh -
$^7$Univ. of S\~ao Paulo, Brazil -
$^8$Wayne State Univ. \\

\begin{abstract} 

Inclusive double differential multiplicities 
and rapidity density distributions of hadrons are presented for 
10.8$\cdot$A~GeV/c Au+Au collisions as measured at the AGS by the E877 
collaboration.
The results indicate that large amounts of stopping and collective
transverse flow effects are present. 
The data are also compared to the results from the lighter Si+Al system. 

\end{abstract} 

\section{INTRODUCTION}

The availability of the first truly heavy beams at the AGS allows
the study of hot and dense nuclear matter over the 
largest volumes that will effectively be available in the laboratory.
Early results on transverse energy production~\cite{AuPRL} 
indicate that with such heavy beams, energy densities predicted 
to lead to a quark-gluon plasma phase transition are probably reached.
The study of hadron spectra provides a more detailed
description of the reaction dynamics and, in particular, of its 
evolution as a function of the mass of the colliding system.

Preliminary results on hadron production at
10.8$\cdot$A$~$GeV/c from the E877 Collaboration were presented at Quark 
Matter 1995~\cite{qm95}.  
Au beam data were taken during the 1993, '94 and '95 AGS heavy ion runs.
Here, we will present mostly '93 data with some preliminary 
results from the '94 run. 
The E877 experimental setup is presented in  
T.~K.~Hemmick's contribution to this conference~\cite{hemmick}. 
It can be summarized into two groups of detectors:
nearly $4\pi$ calorimetry around the target and a forward spectrometer.
The centrality determination is done using the 
calorimeters surrounding the target~\cite{AuMult} and 
the particle spectra discussed here were obtained with the spectrometer.

\section{PROTON DISTRIBUTIONS}

Since at AGS energies creation of baryon-antibaryon pairs is nearly negligible,
the rapidity distribution of protons allows the study of
the energy deposition in the reaction, i.e. the 
nuclear stopping power.

\begin{figure}[thb]
  \epsfig{file=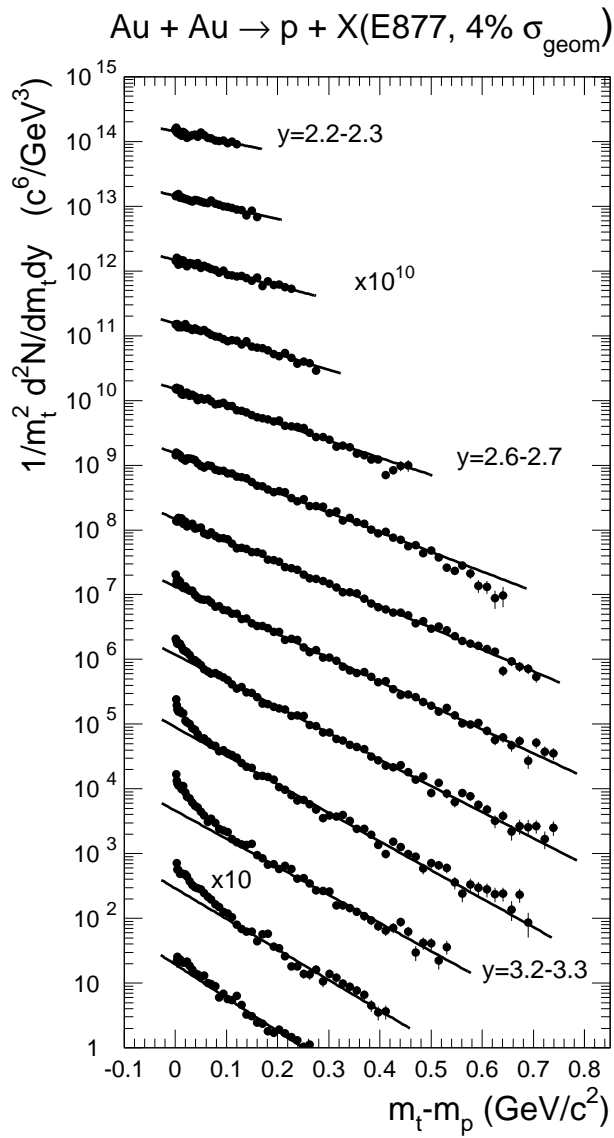,height=6in}
{\begin{minipage}[c]{6cm}
\vspace*{-12.5cm}
  \caption{\small  
Proton transverse mass spectra. 
The data are presented in rapidity bins of 0.1 unit
widths successively multiplied by
increasing powers of 10 for decreasing rapidity.
Errors are statistical. Also shown are exponential fits
which for $y\geq 2.9$ exclude $m_t-m_p<0.2$~GeV/c$^2$.
\label{fig:pmt}}
\end{minipage}}
\end{figure}

E877's measurement of proton transverse mass spectra in Au+Au collisions for
the most central 4\% of $\sigma_{\rm geom}$ 
is presented in 
figure~\ref{fig:pmt}. 
One of the key features of the E877
spectrometer is that its acceptance includes $p_t=0$.
In terms of rapidity, the spectrometer's acceptance for protons 
starts at $y=2.2$ with data extending to $y=3.5$.
The vertical scale in figure~\ref{fig:pmt} is
$1/m_{t}^{2} \cdot dN/dy dm_{t}$ so
that a thermal (Boltzmann type) source would produce an exponential 
distribution.
As in all the figures of this paper, 
the data are presented with statistical error bars only.
 
Overall, the transverse mass spectra exhibit  
thermal shapes with increasing slope parameter when approaching mid-rapidity. 
As expected, deviation from a purely exponential distribution 
is observed around
beam rapidity ($y_{\rm beam}=3.14$) for $m_t-m_p<0.2$~GeV/c$^2$. 
This colder low transverse momentum 
component to the proton spectra can be attributed to spectator
protons from the projectile~\cite{DeeCentral}.

\begin{figure}[thb]
 \begin{center}
  \epsfig{file=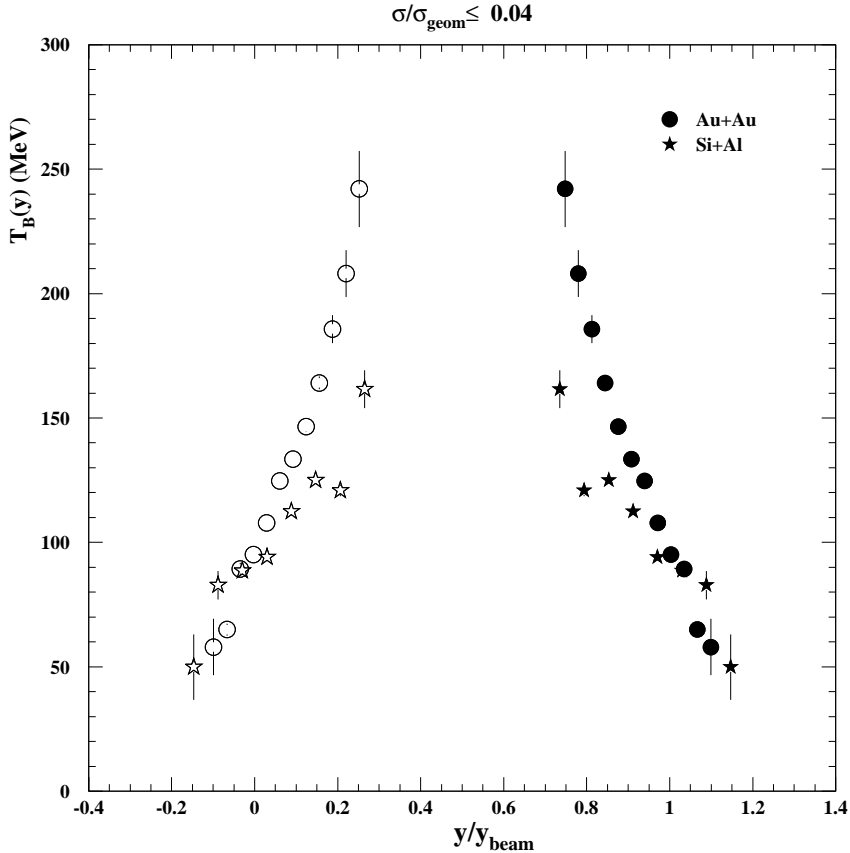,height=4.5in}
 \end{center}
\caption{\small Proton inverse slope constants as a function of rapidity. 
The measured temperature parameters for Au+Au
(filled circles) are compared to data from the 
Si+Al (filled stars) system. 
The data are reflected about mid-rapididity (open symbols).
\label{fig:temps}}
\end{figure}

The values of the inverse slope constants ${\rm T}_{\rm B}$ obtained 
from fitting each of the proton transverse mass spectra with 
an exponential (full lines in figure~\ref{fig:pmt}) are plotted in 
figure~\ref{fig:temps}. 
The data are compared to values obtained for 14.6$\cdot$A~GeV/c 
Si+Al ($y_{\rm beam}=3.44$)~\cite{SiBaryons} and are plotted as a 
function of $y/y_{\rm beam}$.

The measured inverse slope constants for the two systems are quite similar 
around beam rapidity. 
Systematically increasing differences are observed 
between the Au+Au and Si+Al systems which reach their largest values 
closest to the center of mass rapidity.
The present result could be interpreted as being due to
a larger collective transverse flow component in Au+Au compared 
to what was observed in Si+Al~\cite{Expansion}.

\begin{figure}[thb]
 \begin{center}
  \epsfig{file=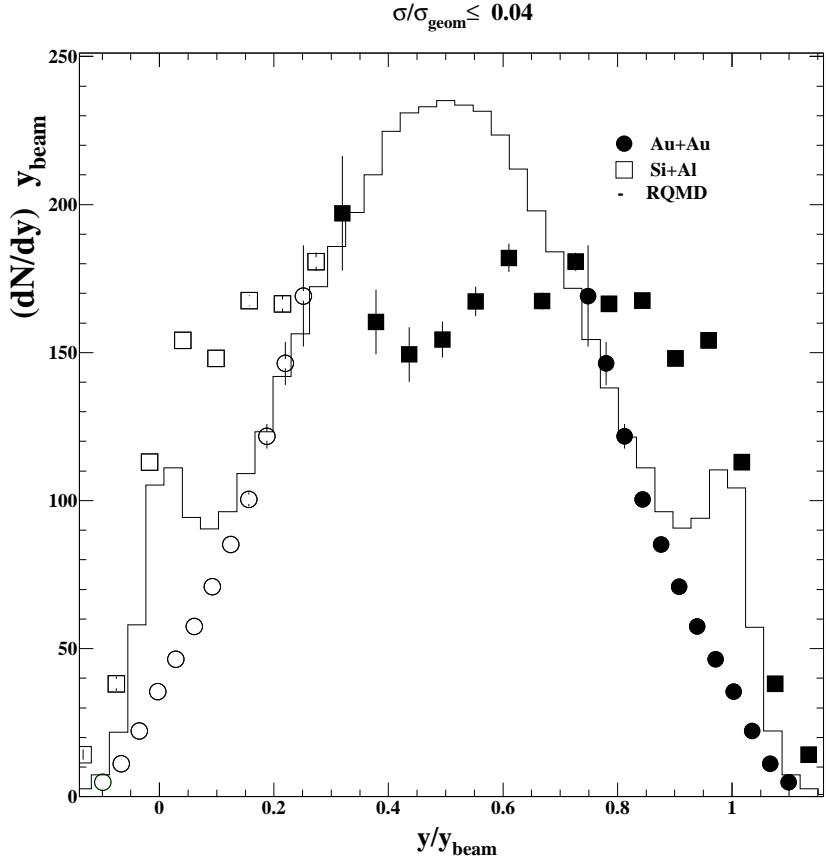,height=4.5in}
 \end{center}
\caption{\small  Rapidity distribution of protons in Au+Au (full 
circles) compared to data from the Si+Al (full stars) system and
to a prediction from RQMD (solid line). The filled symbols are the 
measured points and the open symbols are obtained from reflecting
the measurement about $y_{cm}$.
\label{fig:pdndy}}
\end{figure}

The proton rapidity density distribution shown in
figure~\ref{fig:pdndy} is obtained by integrating the 
transverse mass spectra where data are available and 
extrapolating to $p_t=\infty$ using the Boltzmann fits 
(see figure~\ref{fig:pmt}).
The measured rapidity distribution for Au+Au (circles)
is plotted as a function of $y/y_{\rm beam}$ in order to 
compare with Si+Al data (squares)~\cite{SiBaryons} and is multiplied 
with the proper Jacobian ($y_{\rm beam}$).
The Si+Al rapidity density distribution was also renormalized by 
the ratio of the number of nuleons in Au+Au to that in Si+Al so 
as to better compare the shapes of the two distributions.
The open symbols are obtained by reflecting the experimental 
results about $y_{\rm cm}$. 
The solid line is the result of an RQMD~1.08~\cite{RQMD} calculation 
for Au+Au.

The RQMD calculation exhibits a structure near $y_{beam}$ originating from 
projectile protons which have had 
no or minimal interaction with the target.
This structure includes protons originating from the dissociation of
the spectators treated as unbound in RQMD.
The measured rapidity distribution of protons in Au+Au 
is significantly narrower than what was observed for Si+Al.
The distribution is still wider than predicted by an 
isotropic thermal distribution, using the temperature 
deduced from the pion spectra, and is
consistent with a longitudinaly expanding thermalized 
source~\cite{stachel}.
Although the stopping was already shown to be high in Si+Al~\cite{SiStopping}, 
the measured Au+Au rapidity distribution indicates that an even
larger amount of stopping is involved. 
This is understood as being a consequence of the smaller
surface to volume ratio and increased average number of rescatterings 
in the Au+Au system. 
This interpretation is supported by the RQMD calculation
that reproduces both overall widths of the 
Si+Al~\cite{SiBaryons} and the Au+Au measurements.

\section{PION DISTRIBUTIONS} 

Pions are produced copiously at AGS energies. Because of their
light masses and large cross sections for interaction in nuclear matter
they are expected to thermalize easily. 
Furthermore, their spectra are not as affected as
those of heavier particles by a given collective flow velocity and thus  
are good probes for studying thermal properties at freeze-out.

\begin{figure}[thb]
 \begin{center}
  \epsfig{file=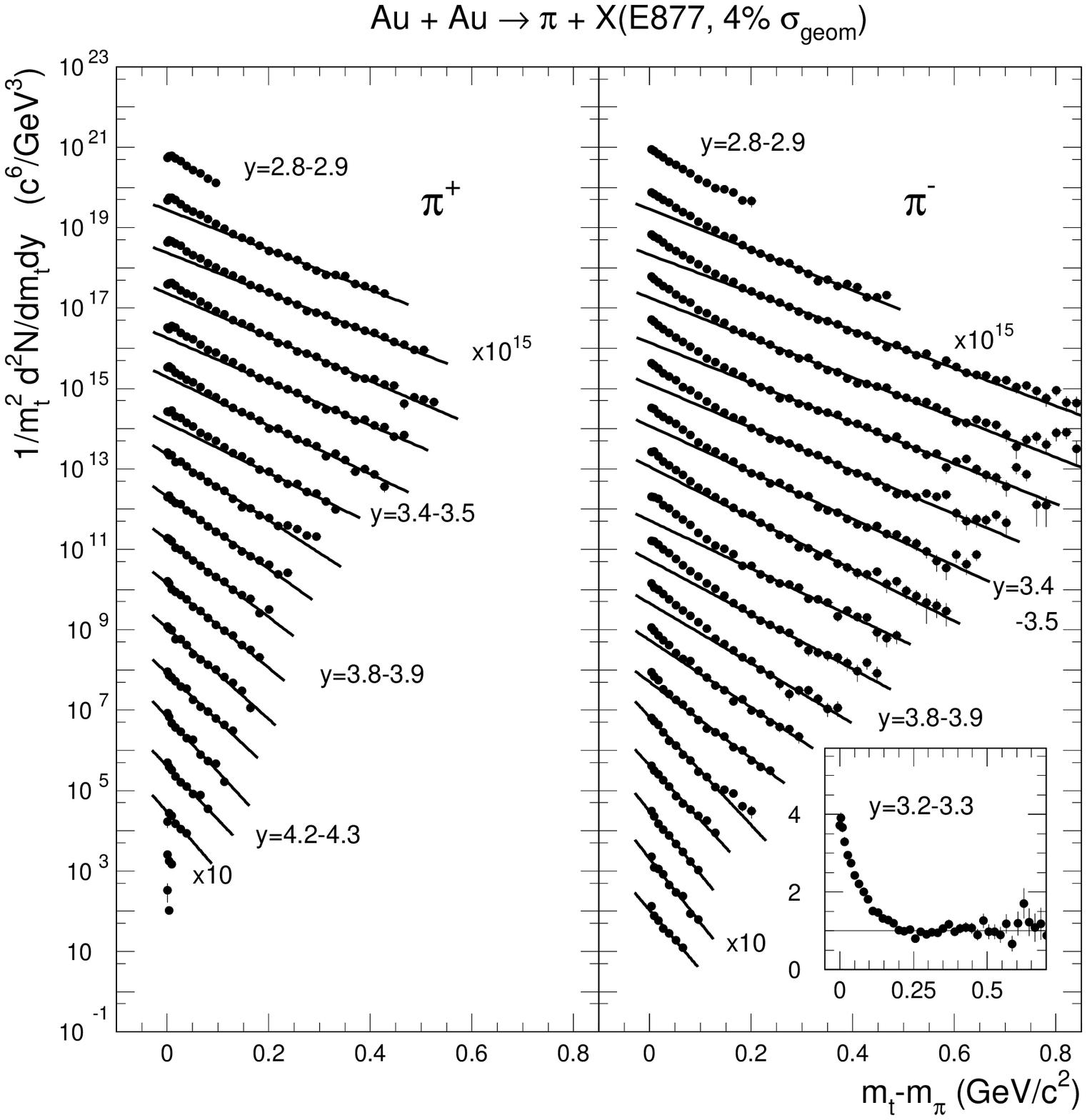,height=5.55in}
 \end{center}
\caption{\small  
Pion transverse mass spectra.
The data are presented in rapidity bins of 0.1 unit widths successively 
multiplied by increasing powers of 10.
\label{fig:pimt}}
\end{figure}

The pion spectra in Au+Au collisions for the most central 4\% of
$\sigma_{\rm geom}$ are shown in 
figure~\ref{fig:pimt} as a function of transverse mass and rapidity.
The pion data extend to 0.8~GeV/c$^2$ in $m_t-m_{\pi}$
and cover rapidities 2.8 to 4.5.
The transverse mass spectra are exponential with slopes
decreasing with increasing rapidity.
A clear enhancement above a pure exponential is observed for
$m_t-m_\pi < $~0.2~GeV/c$^2$.
A similar effect was already observed for the Si+Pb system
and was explained by the contribution from decay pions 
of $\Delta$ resonances~\cite{Delta}.
To better display the enhancement for the $y=3.2-3.3$ rapidity slice, 
the data were divided by an exponential fitted to the part of
the spectra above $m_t-m_{\pi} >$~0.2~GeV/c$^2$.
As shown on the insert, the enhancement reaches values as large as 4 at 
$m_t-m_{\pi} =$~0~GeV/c$^2$. 
The enhancement is systematically increasing as one goes toward $y_{\rm cm}$. 

\begin{figure}[thb]
 \begin{center}
  \epsfig{file=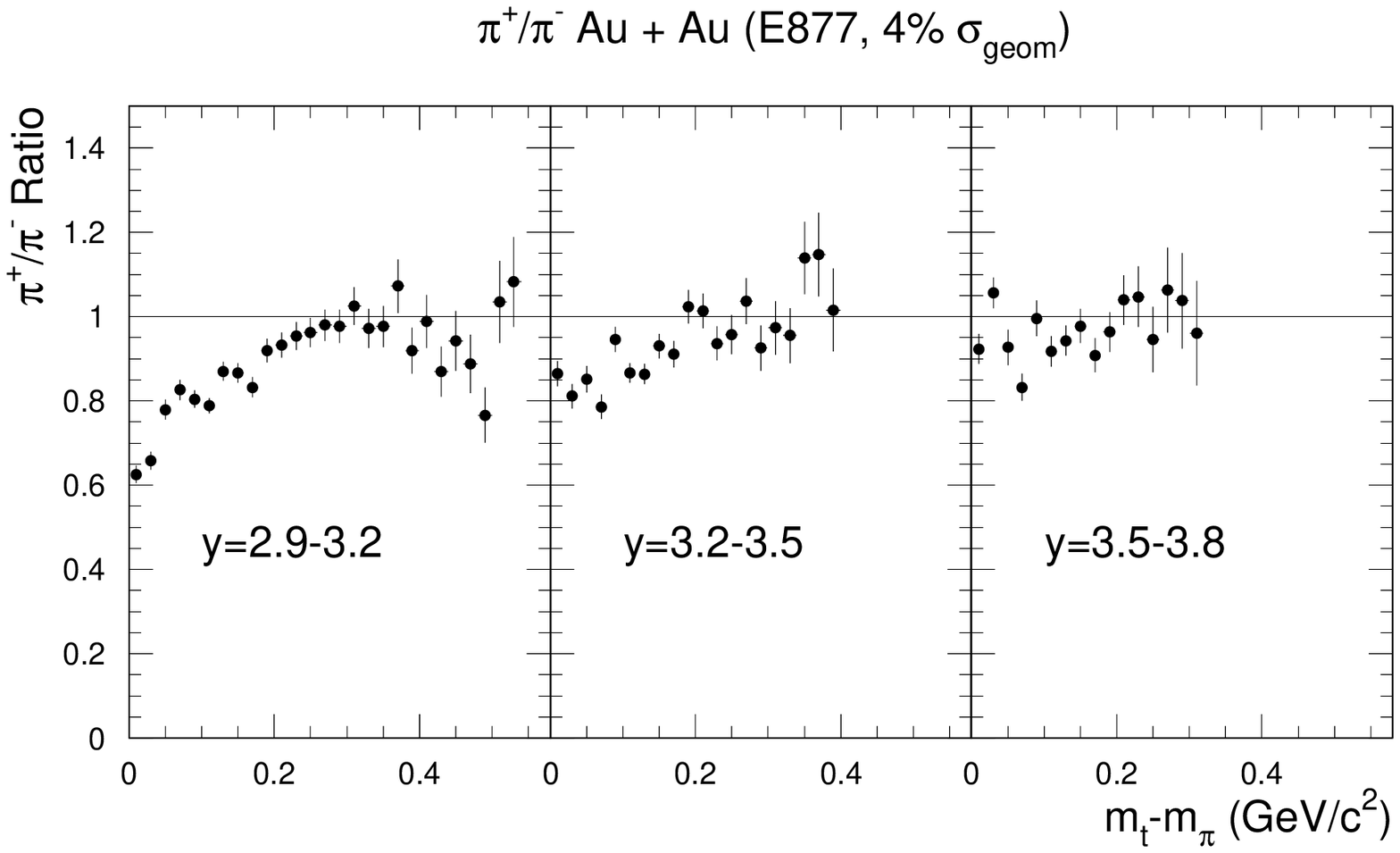,width=5in}
 \end{center}
 \caption{\small Ratio of the transverse mass spectra of $\pi^+$ and $\pi^-$ 
in 3 rapidity bins.
\label{fig:piratio}}
\end{figure}

Close inspection of the transverse mass curves of
figure~\ref{fig:pimt} reveals that the deviations from a pure 
exponential emission are systematically larger for
$\pi^{-}$ than for $\pi^{+}$.   
The charge asymmetry of the pion production is better studied by
plotting  the ratio of $\pi^{+}$ to $\pi^{-}$ as done 
in figure~\ref{fig:piratio}.  
In order to improve statistical errors 
in the ratio, the data were divided into three large rapidity bins.
The $\pi^{+}/\pi^{-}$ ratio is consistent with unity at large
transverse mass values. 
A strong charge asymmetry is observed starting at 
$m_t-m_{\pi}< 0.2$~GeV/c$^2$ with a minimum value of
$\pi^{+}/\pi^{-}\simeq 0.6$ at $m_t-m_{\pi}=0$~GeV/c$^2$ for the
$y=2.9-3.2$ rapidity slice. 
The asymmetry is also observed to systematically decrease 
as a function of rapidity.
The observed rapidity dependence is in line with the  
charge asymmetry of $\pi^{+}/\pi^{-} \simeq 0.5$ measured
by the E866 collaboration near 
$y_{\rm cm}$ at similar centrality~\cite{qm95e866}. 

The observed pion charge asymmetry and its rapidity dependence 
could be due to the different Coulomb
potentials seen by each charge type at freeze-out.
Coulomb interactions have been discussed
at this conference in terms of their influence on the interpretation 
of particle correlations~\cite{dariusz,kadija,baym}. 
The study of the Coulomb effect on the shape of the particle spectra
will be pursued since it provides a different approach to the 
determination of the spacetime particle distribution at 
freeze-out.

\begin{figure}[thb]
 \begin{center}
  \epsfig{file=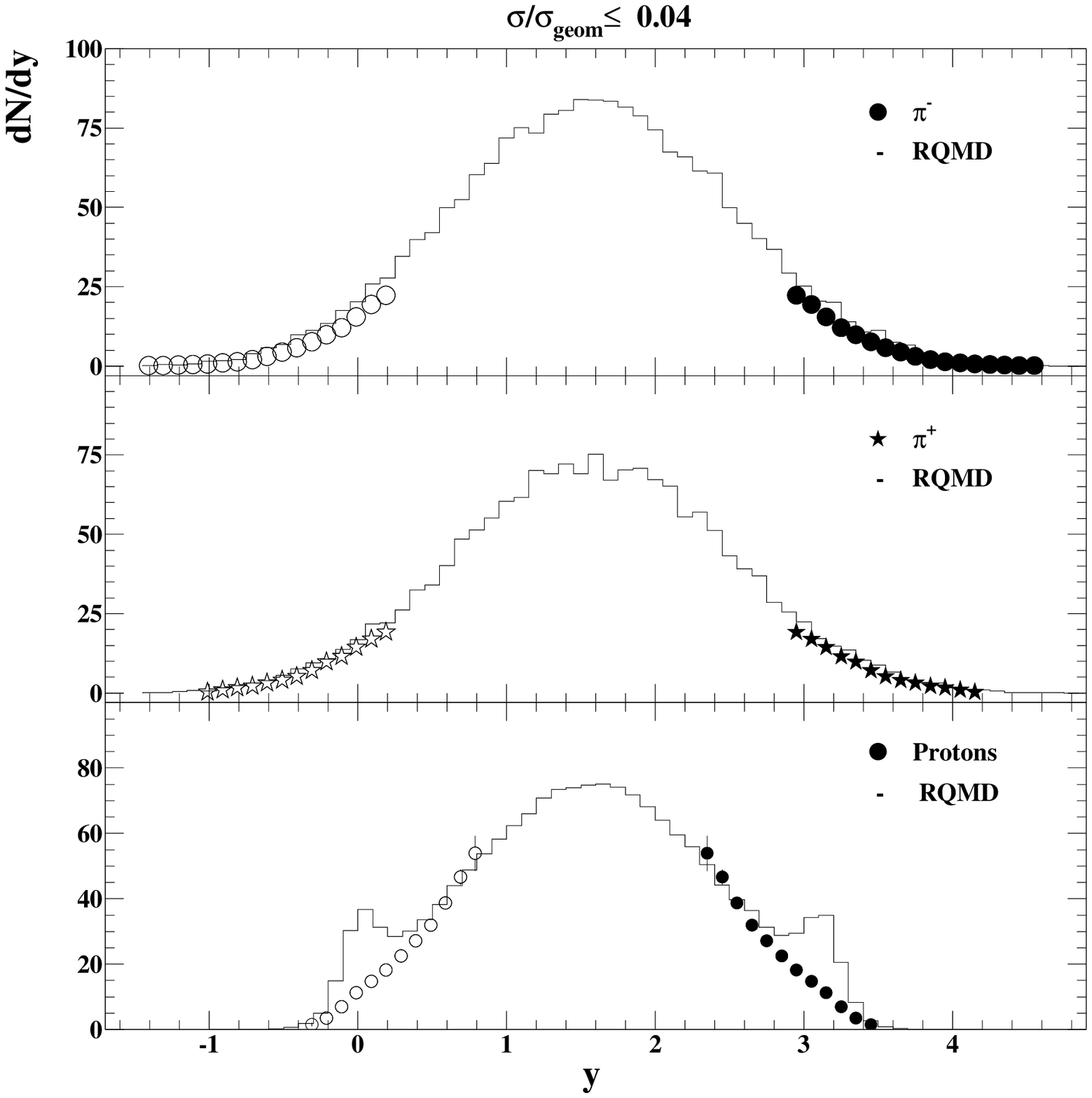,height=5in}
 \end{center}
 \caption{\small Rapidity distributions for $\pi^{+}$, $\pi^{-}$ and
protons.  The data (solid symbols) are reflected about $y_{\rm cm}$ 
(open symbols). An RQMD calculation is overlaid (solid line) 
for comparison.
\label{fig:pidndy}}
\end{figure}

The rapidity distribution for pions obtained by integrating
the transverse mass spectra is shown in figure~\ref{fig:pidndy}.
The distributions are plotted as a function of laboratory rapidity,
covering the rapidity range $2.9<y<4.4$. 
The RQMD calculations agree very well with our data over their
entire range.
A similar agreement was also observed for the the Si+Pb 
system~\cite{qm93delta}. 

The proton rapidity distribution is also included at the bottom
of figure~\ref{fig:pidndy}.
Contrary to what was observed in the Si+Al~\cite{Expansion},
where the proton distribution was wider than that of the pion,
the rapidity distribution widths are similar for Au+Au.
The measured narrowing of the proton distribution from the Si+Al is 
reproduced by the RQMD calculation.
These observations are consistent with full stopping and 
similar values of longitudinal and transverse expansion
of $\beta\sim0.5$~\cite{stachel}.

\begin{figure}[htb]
  \epsfig{file=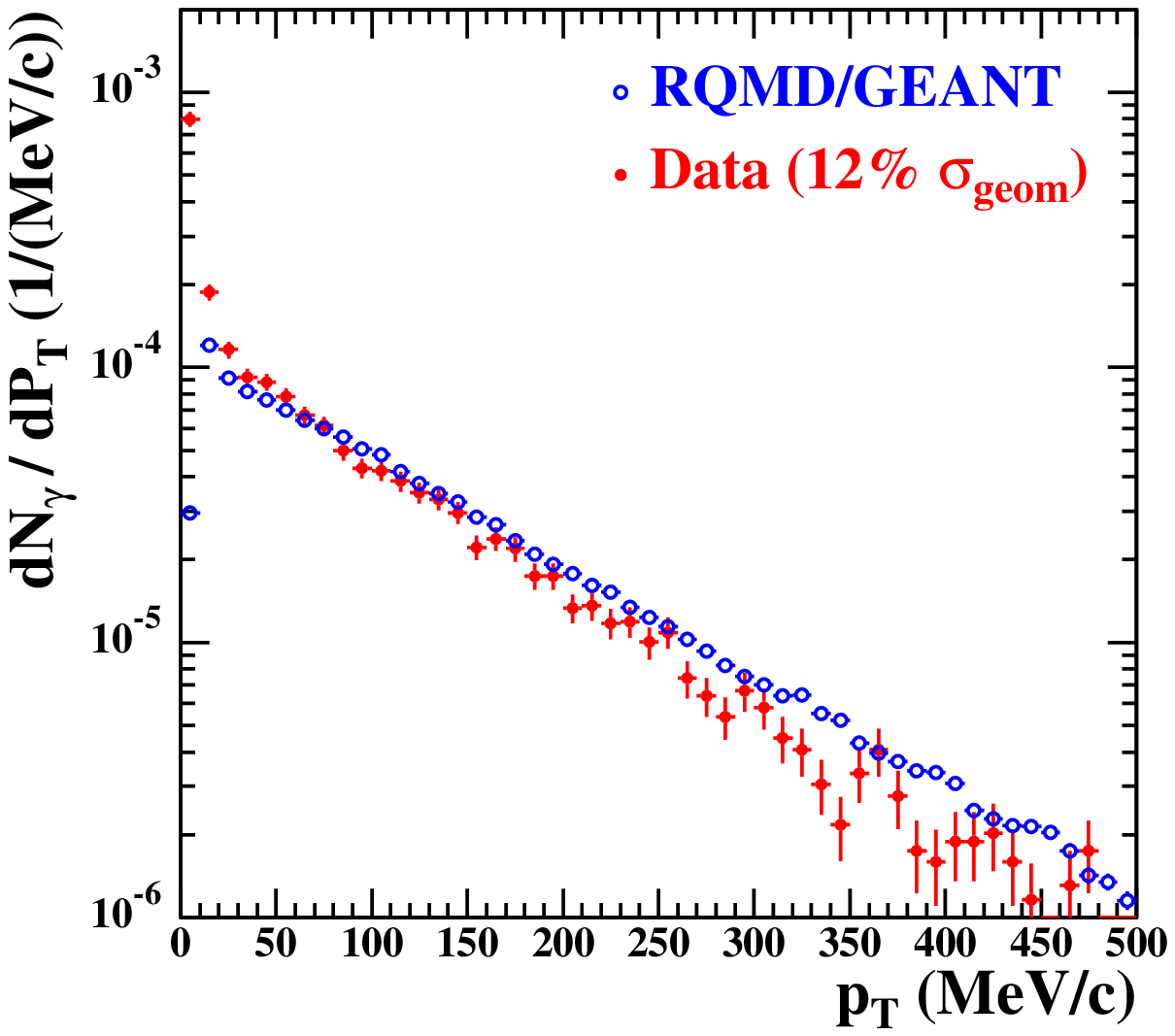,height=3.in}
  \epsfig{file=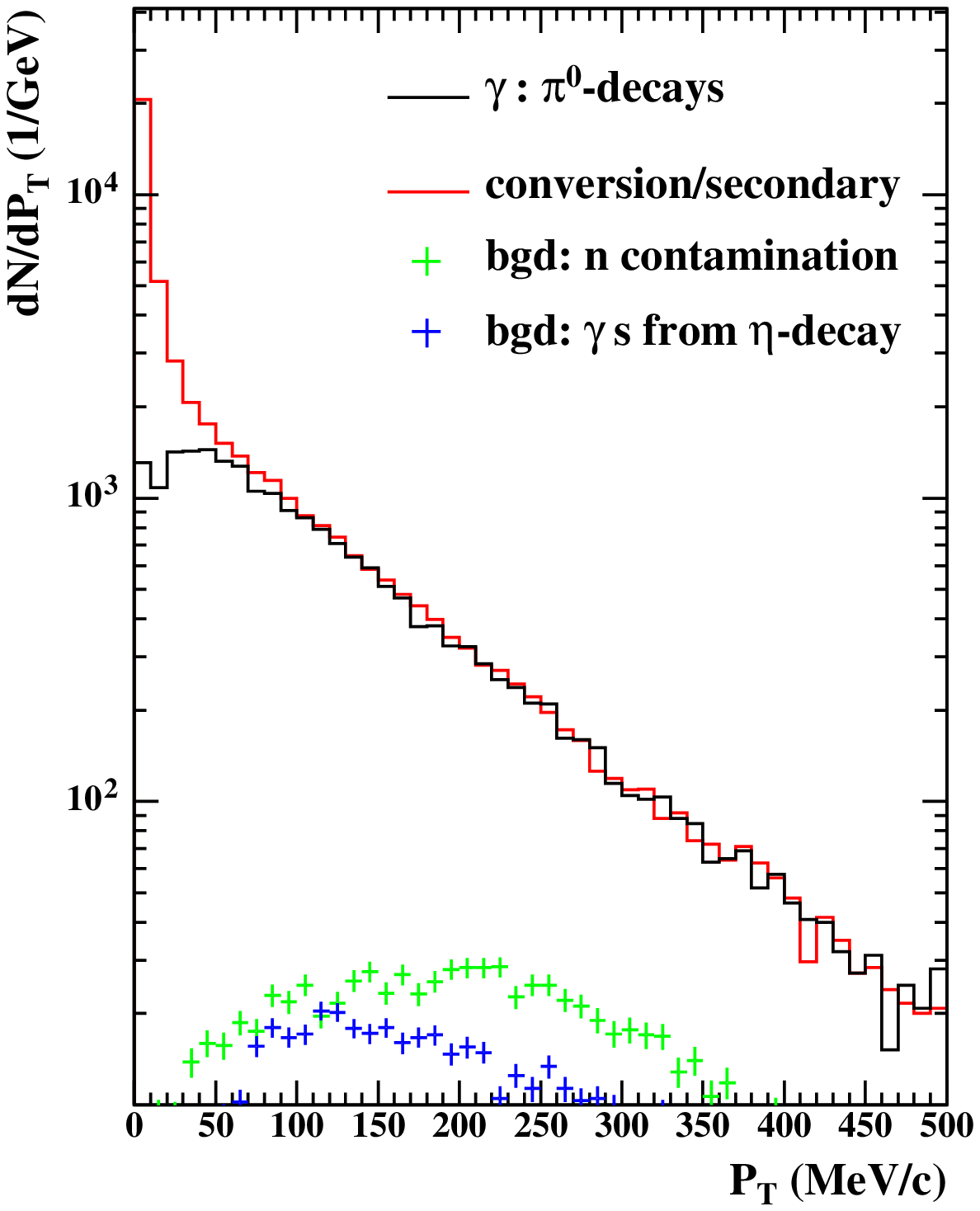,height=2.95in}
 \caption{\small Photon measurement made with the CsI detector. Shown
on the left is the data and an RQMD+GEANT calculation and on the right 
the decomposition of the different sources that contribute to the
calculation.
\label{fig:pizero}}
\end{figure}

The inclusive photon spectra, and indirectly the $\pi^{\circ}$ yield,
was measured using a CsI photon detector~\cite{zou}. 
This detector was placed 35 m from the target at an angle of 
$\theta = 4.8^{\circ}$ and covered $\delta\theta \sim 0.34^{\circ}$. 
The photon inclusive transverse momentum measurement 
is shown on left side of figure~\ref{fig:pizero} (open symbols). 
Because of the small size of the detector, reconstruction of the 
$\pi^{\circ}$ using the two decay photons was impossible. 
A model dependent method was used to derive a limit on $\pi^{\circ}$ 
production using single photons.
An RQMD calculation was performed followed by 
tracking of the decay photons through the E877 
spectrometer using a GEANT (full symbols)~\cite{GEANT}. 
The RQMD+GEANT calculation is in fair agreement with the 
shape of the measured photon spectra except for the lowest 
transverse momentum values $p_t<100$~MeV/c.

As the calculation shows on the right side of figure~\ref{fig:pizero}, 
the single photon spectrum is dominated, above $p_t > 100$~MeV/c,  
by photons that originate from $\pi^{\circ}$ decays.
The calculation also indicates that the dominant contributions 
at low transverse momentum ($p_t< 100$~MeV/c) are secondary photons.
These secondary photons originate from interactions in the 
spectrometer material located upstream of the CsI detector.
The contributions to the measured photon spectrum from neutron
contamination and $\eta$ decays are also shown.
A careful inspection of the higher end of the inclusive photon 
spectrum on left side of figure~\ref{fig:pizero} reveals that 
the calculation systematically overpredicts the yield of photons. 
Using the photon spectrum decomposition made above, the discrepancy
is found to correspond to an overprediction by RQMD 
of roughly $\sim 20\%$ of the $\pi^{\circ}$ yield~\cite{zou}. 
RQMD predicts a ratio between the pion yields of 
$\pi^+/\pi^{\circ}/\pi^- = 1/1.26/1.16$ at a centrality of 
12\% of $\sigma_{\rm geom}$.  
Thus the present result does not support the predicted excess 
of neutral pions.

\section{KAON DISTRIBUTIONS}

\begin{figure}[htb]
 \begin{center}
  \epsfig{file=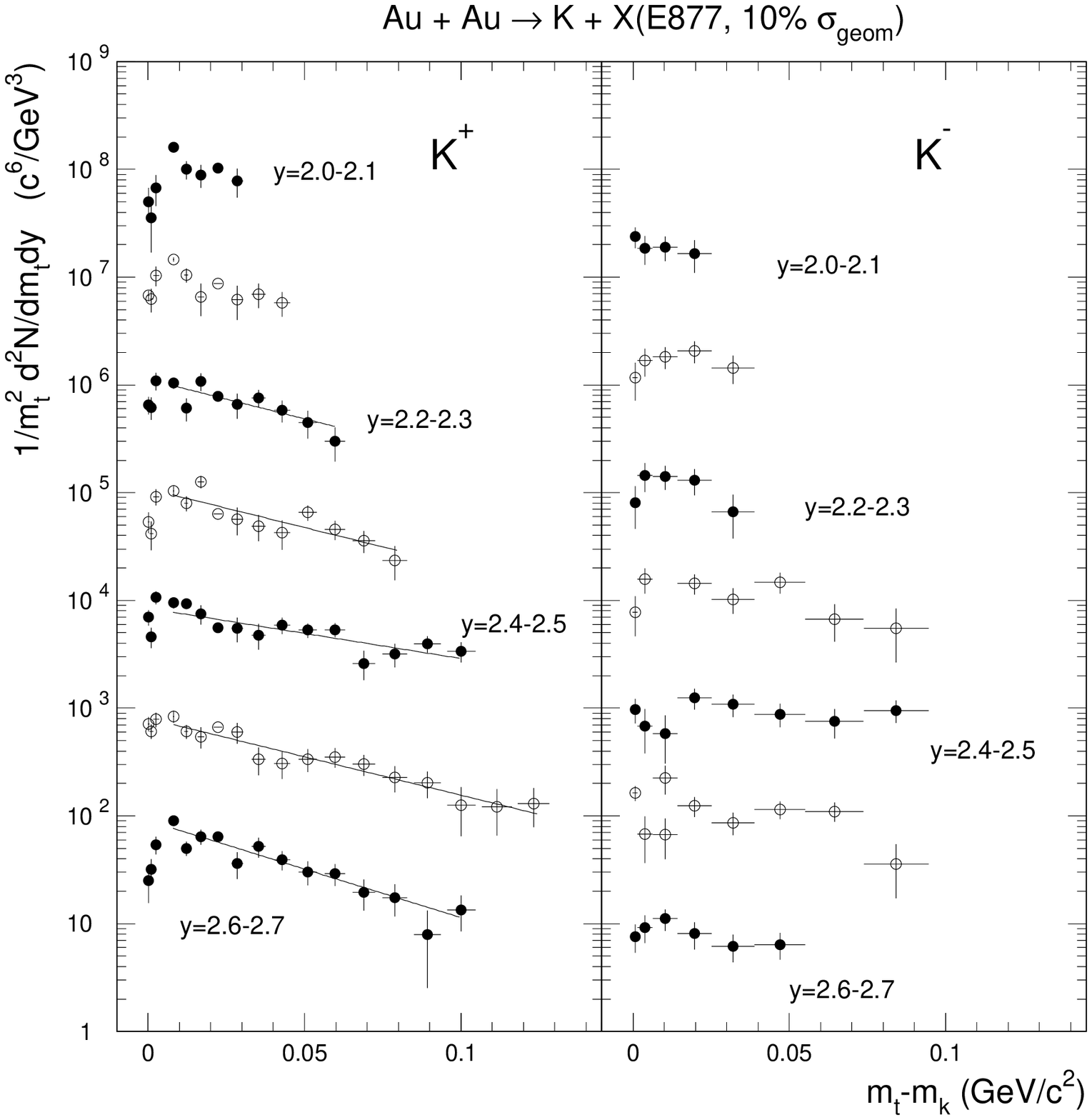,height=5in}
 \end{center}
\caption{\small  Transverse mass spectra for K$^+$ on the left
and K$^-$ on the right. The data are presented in rapidity bins 
of 0.1 unit widths successively multiplied by increasing powers of 10.
\label{fig:kaons}}
\end{figure}

The study of the production kaons and other strange particles 
provides information on the level of chemical equilibrium achieved and
is also relevant to determine whether chiral restoration is achieved.
At Quark Matter 95, preliminary kaon spectra were presented 
that created a high level of interest~\cite{qm95}.
The measured transverse mass spectra of K$^+$ seemed to exhibit
an unexpected enhancement at the lowest transverse mass values. 

In order to validate the shape of these spectra, new measurements of the
kaon spectra were performed in 1994 with slightly 
different experimental conditions.
In addition, the preliminary results from the 1993 run were complemented
by an independent re-analysis of the data~\cite{thesis}.

The results from the re-analysis of the 1993 data for the 
top 10\% of $\sigma_{\rm
geom}$ are plotted in figure~\ref{fig:kaons}.
The $m_t$ scales are divided into bins equal to 20~MeV/c in
$p_t$ for K$^+$ and 40~MeV/c for K$^-$.
The K$^+$ spectra no longer show a statistically significant 
enhancement at low transverse mass.
The origin of the structure was traced to an albedo source
located at the spectrometer's collimator edge, closest 
to the beam trajectory.
The present results are 
confirmed by the preliminary analysis of the 1994 data.
Both sets of data show an indication 
of a dip at very low $p_t$ in the K$^+$ spectra.

The fitted exponential inverse slopes have values of 60 to 90~MeV which 
are consistent with those presented at Quark Matter '95 for K$^+$.
These values are lower than the ones obtained from the 
E866 data (T$_{\rm B}\sim 150$ to $170$~MeV) in the same 
rapidity range~\cite{stachel,akiba}.
The E866 data cover transverse mass values of up to 1~GeV/c$^2$
whereas our measurement covers the first 0.1~GeV/c$^2$. 

\section{DEUTERON DISTRIBUTIONS}

\begin{figure}[th]
 \begin{center}
  \epsfig{file=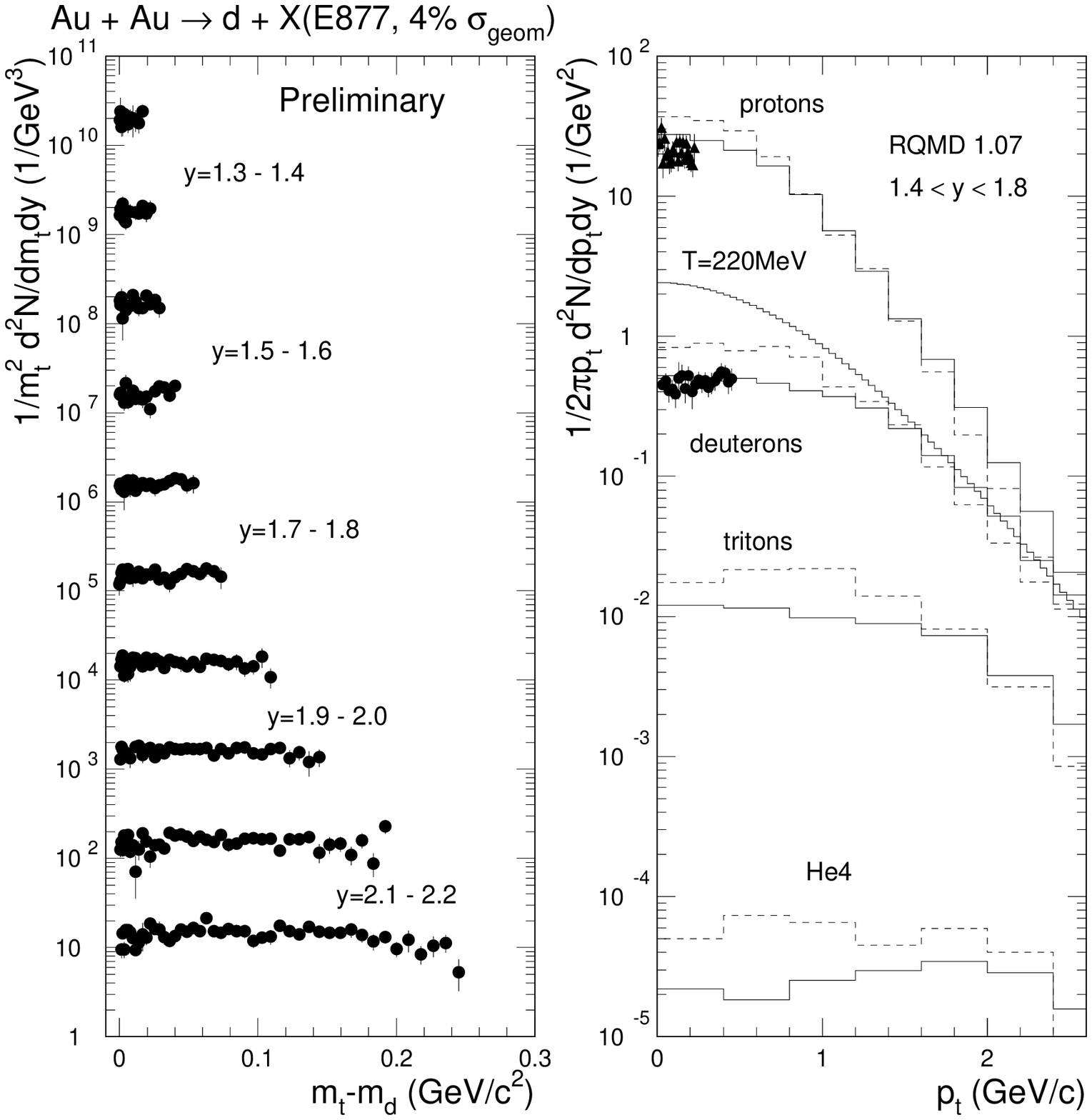,height=5.5in}
 \end{center}
\caption{\small  Transverse mass spectra for deuterons are shown on the
left. The data are presented in rapidity bins of 0.1 widths successively
multiplied by increasing powers of 10. Different RQMD predictions for the
rapidity bin $1.4<y<1.8$ are compared to deuteron (circles) and 
proton (triangles) data on the right.
The dashed histograms correspond to predictions of standard RQMD while
the full histograms include the effect of mean field. 
The full line is a Boltzmann distribution fitted to the large $p_t$
region of the RQMD result for deuterons.
\label{fig:deuts}}
\end{figure}

The production of composite particles at the AGS has yet to be
studied extensively.
Their production is just starting to be considered in the
cascade models. 
Deuterons are, because of their mass, good probes to study 
expansion and collective transverse flow velocities.

Preliminary results from the measurement of deuteron production from the
1994 run are presented in figure~\ref{fig:deuts} for 
the most central collisions (4\% of $\sigma_{\rm geom}$). 
The deuteron acceptance covers the rapidity 1.2 to 2.2
and the data extend to 0.25~GeV/c$^2$ in $m_t-m_d$ for $y=2.1-2.2$.  
The measured deuteron spectra are very flat over the covered 
transverse mass range and exhibit only a slight rapidity dependence.

On the right side of figure~\ref{fig:deuts},  
data for deuterons (circles) and protons (triangles)
are compared to the corresponding 
RQMD calculations (lines) for the production
of different types of composites (p, d, t, He$^4$) 
in the rapidity slice $1.4<y<1.8$~\cite{mattiello}.
The result from the cascade version of RQMD 
(dashed histogram) is found to overshoot the data. 
A modified version of RQMD that simulates the presence of a mean field
(full line histogram) reproduces the measured deuterons yield
quite well but still overshoots the proton data. 
A Boltzmann fit to the large $p_t$ region of the RQMD deuteron 
calculation (full line) shows that even the 
simple cascade model deviates from a thermal shape 
and produces much flatter distributions at low $p_t$.

The spectral shapes of the deuteron data support the importance of 
the mean field to describe the measured yields.
The model predicts an increasing flattening with mass and thus  
the measurement of the spectra of more complex composites 
would serve to better determine the importance of mean field effects in 
relativistic heavy ions collisions. 

\section{CONCLUSION}
 
We have presented new data on the particle distributions of hadrons
produced in Au+Au collisions at AGS energies. 
The rapidity distributions show increased stopping relative 
to lighter systems leading to a center of mass region richer in baryons. 
The rapidity dependence of the slopes of the particle spectra 
and the rapidity distributions show evidence of large collective 
transverse and longitudinal flow. 
The data are consistent with a larger collective component in the 
Au+Au final state than what was observed for the Si+Al system. 
The slope and yield of the measured deuteron spectra at low $p_t$
suggest the presence of sizable mean field effects.
The addition of a new vertex detector system to the E877 
spectrometer and the better statistical samples from the 
runs of 1994 and 1995 should soon provide interesting new 
results, particularly on deuteron, K$^-$ 
and $\Lambda$ production.

The avalaible data on both global observables and particle spectra 
in Au+Au collisions are consistent with the formation of
baryon rich nuclear matter at density and temperature close
to that expected for a phase transition. 
It is particularly interesting to note that heavy systems 
provide us with new observables, such as collective flow,  
mean field and Coulomb effects, that will help us to better
understand the space-time evolution of the system during the
collision and yield additional signatures for new phenomena.

Support from US DoE, the NSF, the Canadian NSERC, and CNPq Brazil 
is gratefully acknowledged.


\end{document}